# Assessing the effectiveness of test-trace-isolate interventions using a multi-layered temporal network

Yunyi Cai[1], Weiyi Wang[1], Lanlan Yu[1], Ruixiang Wang[1], Gui-Quan Sun[2,3], Allisandra G. Kummer[4], Paulo C. Ventura[4], Jiancheng Lv[1], Marco Ajelli[4,#], Quan-Hui Liu[1,#,*]

1 College of Computer Science, Sichuan University, Chengdu, China
2 Department of Mathematics, North University of China, Taiyuan, China
3 Complex Systems Research Center, Shanxi University, Taiyuan, China
4 Laboratory for Computational Epidemiology and Public Health, Department of Epidemiology and Biostatistics, School of Public Health, Indiana University Bloomington, Bloomington, Indiana, United States of America

[#] Senior author
[*] Corresponding author

**Abstract**

In the early stage of an infectious disease outbreak, public health strategies tend to gravitate towards non-pharmaceutical interventions (NPIs) given the time required to develop targeted treatments and vaccines. One of the most common NPIs is Test-Trace-Isolate (TTI). One of the factors determining the effectiveness of TTI is the ability to identify contacts of infected individuals. In this study, we propose a multi-layer temporal contact network to model transmission dynamics and assess the impact of different TTI implementations, using SARS-CoV-2 as a case study. The model was used to evaluate TTI effectiveness both in containing an outbreak and mitigating the impact of an epidemic. We estimated that a TTI strategy based on home isolation and testing of both primary and secondary contacts can contain outbreaks only when the reproduction number is up to 1.3, at which the epidemic prevention potential is 88.2% (95% CI: 87.9%-88.5%). On the other hand, for higher value of the reproduction number, TTI is estimated to noticeably mitigate disease burden but at high social costs (e.g., over a month in isolation/quarantine per person for reproduction numbers of 1.7 or higher). We estimated that strategies considering quarantine of contacts have a larger epidemic prevention potential than strategies that either avoid tracing contacts or require contacts to be tested before isolation. Combining TTI with other social distancing measures can improve the likelihood of successfully containing an outbreak but the estimated epidemic prevention potential remains lower than 50% for reproduction numbers higher than 2.1. In conclusion, our model-based evaluation highlights the challenges of relying on TTIs to contain an outbreak of a novel pathogen with characteristics similar to SARS-CoV-2, and that the estimated effectiveness of TTI depend on the way contact patterns are modeled, supporting the relevance of obtaining comprehensive data on human social interactions to improve preparedness.

**Background**

Public health strategies to prevent and control the spread of infectious diseases are crucial for reducing disease burden (1-5). In the early stage of newly emerging infectious diseases, public health strategies gravitate towards non-pharmaceutical interventions (NPIs) given the time required to develop targeted treatments and vaccines. One of the most commonly adopted NPIs is Test-Trace-Isolate (TTI), which plays a key role in both mitigating and understanding the transmission dynamics of an emerging infectious disease (6-8). TTI is based on the identification of infected individuals, often triggered by the presence of symptoms, who are then tested, isolated (e.g., in their place of residence or in dedicated facilities), and their contacts are traced. If contacts are found to be positive, they are isolated as well, and the TTI process is repeated for their contacts. In general, TTI tends to be very effective for diseases with high proportion of symptomatic infections and little proportion of pre-symptomatic transmission (e.g., Ebola, MERS, and SARS-CoV-1) (9-14).

During the early phase of the COVID-19 pandemic, several countries adopted TTIs to try to contain or at least to mitigate the spread of SARS-CoV-2 (15-17). However, the extent to which TTIs were effective is still debated (6, 7, 18-20). This is partially due to the widely different implementation of TTI in different countries at different time points in the pandemic and heterogeneous levels of population adherence. For example, during the early stage of COVID-19 in China, both the primary and the secondary contacts of the confirmed cases were traced and quarantined for 14 days in the dedicated facilities (21). In contrast, only household contacts of confirmed cases were traced and required to self-isolate at home in the United States (22). Moreover, TTI was generally implemented in tandem with a set of other NPIs, which makes it hard to disentangle the effectiveness of TTI alone (23-25). To overcome this limitation, mathematical modeling studies have been performed where TTI can be tested alone or in conjunction with other strategies. However, this did not resolve the debate as model-based evaluations of TTI effectiveness have yielded highly variable results. For example, Hellewell et al. developed a branching process model of SARS-CoV-2 transmission showing that if 70% of contacts are traced, the majority of outbreaks can be controlled (26). Kerr et al. used an open-source agent-based model leveraging detailed demographic and epidemiological data for the Seattle region, USA, and showed that high but achievable levels of TTIs are sufficient to curtail SARS-CoV-2 spread (18). Kucharski et al. (7) and Chiu et al.(27) showed that increased testing and contact-tracing capacity are paramount for mitigating the spread of SARS-CoV-2. On the other hand, Gardner et al. (28) used compartmental models to show that contact tracing could reduce SARS-CoV-2 reproduction number by no more than 20%. Davis et al. (29) developed a branching process model that showed that even well-implemented contact tracing can provide at most a 15% reduction of SARS-CoV-2 reproduction number. Wang et al. (30) and Contreras et al. (6) showed that TTI alone is insufficient to neither end nor contain a COVID-19 outbreak.

Most of the aforementioned discrepancies can be explained by the assumptions made by different modeling studies. For example, one of the budling blocks of TTI is the identification of contacts. As such, the contact network used to model transmission plays a central role. Some studies have assumed homogeneous mixed patterns, failing to account for the heterogeneities in human contact patterns (7, 19, 26, 29, 31). Other studies have developed complex contact networks but did not incorporate heterogeneities in the risk of infection by social setting (e.g., been members of the household entail a different risk of infection that a short-lasting contact at a grocery store) (32, 33).

To address these gaps, previous studies have proposed multi-layer contact networks, categorizing different social settings into different layers. For example, Liu et al. (34) developed a model based on a multiplex network consisting of household, school, and community layers. Similarly, Kerr et al. (35) used an agent-based model based on contact surveys to construct a multi-layered synthetic population network, including households, schools, workplaces, long-term care facilities, and communities, to project epidemic trends, explore intervention scenarios, and estimate resource needs. Zhang et al. (36) further refined the modeling of contact patterns by incorporating temporal information into a multi-layer contact network that allows contacts between individuals to change dynamically as the simulated epidemic progresses.

This study aims to provide a comprehensive evaluation of the effectiveness of TTIs. To this aim, we proposed a multi-layer temporal network based on the synthetic population to model the contact patterns in the population and their temporal changes. Then, we simulate the spread of a respiratory pathogen on this network. As a case study of a respiratory pathogen, we use the ancestral SARS-CoV-2 lineage. Finally, we simulated seven alternative implementations of TTI to reflect different strategies that were implemented by different countries during different periods of the COVID-19 pandemic. We simulated the model under multiple scenarios and evaluated the mitigation effects of TTI alone or in combination with other NPIs.

## Methods
### Synthetic population
We produced a synthetic population of China by leveraging highly detailed micro-level household survey (37) and macro-level demographic statistics data (38, 39). The household survey includes the age of each member in the surveyed households, while the demographic statistics includes the school enrollment rate, employment rate, and household, school, and workplace size distributions. The synthetic population consists of synthetic individuals that are grouped into households, schools, workplaces and the community, corresponding to the main settings for SARS-CoV-2 transmission (34).

To reconstruct households in the synthetic population, we use a bootstrap sampling method (40), where we: 1) randomly sample the household size from the distribution of household sizes in China; 2) randomly select one household with the sampled size from the household survey data. Households are sampled until the synthetic population includes ~0.5 million individuals. Ages of each household member correspond to those from the sampled household. For each individual in the household, we assign synthetic individuals as students, workers, or other (e.g., retired, preschool children, unemployed) based on the school enrollment and employment rates. Schools and workplaces are then generated based on the size distribution of schools and workplaces, respectively. We assumed all individuals in the synthetic population are present in the community. A schematic representation of the synthetic population is shown in Fig. 1A. Details on the generation of the synthetic population are reported in Additional File 1.

### Multi-layer temporal contact network
The synthetic population defines social groups such as households, schools, and workplaces by matching the demographic characteristics of the analyzed country. To model the heterogeneous and temporal characteristics of human contact patterns within these groups (namely, schools, workplaces, and the community), we relied on the distribution of the number of contacts from a published contact survey conducted in Shanghai, China, between December 2017 and May 2018(41).

At each time step of the simulation, all members of the same household are assumed to be connected to each other. For the school and workplace layers, we used a configuration model (42) to construct the contact network for each school and workplace in their respective layer at each time step. Specifically, for each infectious individual $i$, we randomly sample a number $m$ from the layer-specific distribution of the number of contacts in the school/workplace. Then, we select $m$ members from the same school/workplace to be the contacts for infectious individual $i$ in that time step. As individuals interact with some classmates/colleagues more frequently than with others (43), a fraction $q$ of the contacts for individual $i$ are sampled from their recurring contacts. Constructing the community layer is done using a similar process; however, the contacts for each individual $i$ are randomly selected from the entire synthetic population without considering recurring contact. For comparison with the temporal network, we randomly chose a snapshot from the temporal network as a static network. A schematic visualization of multi-layer temporal contact network is shown in Fig. 1B and details on the construction of the temporal network are described in Additional File 1.

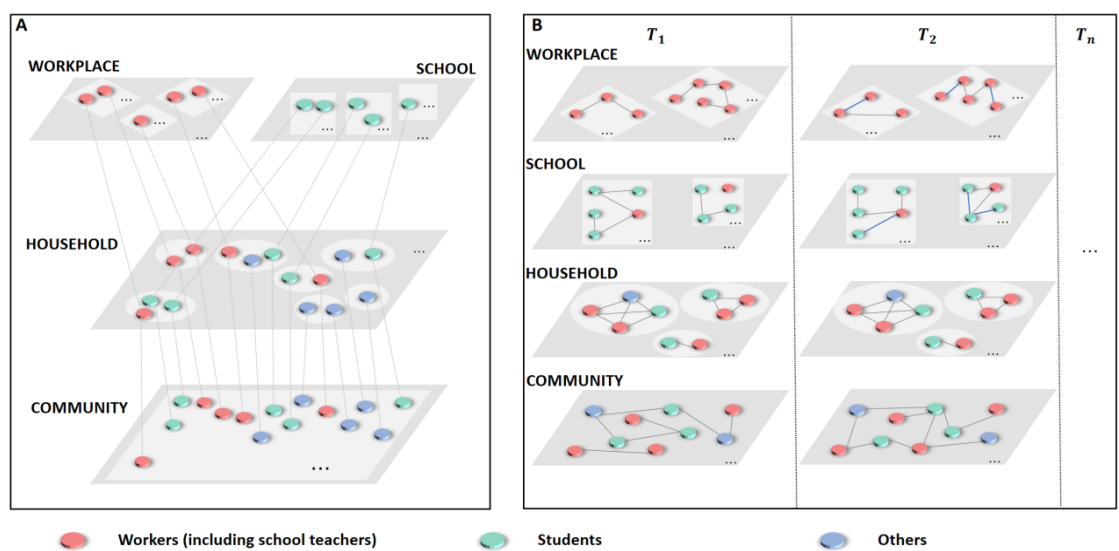

**Fig. 1 Schematic representation of the synthetic population and of the temporal contact network.** **A** Schematic representation of how synthetic individuals are connected between the workplace, household, school, and community layers. **B** Schematic representation of how synthetic individuals are connected within each layer.

**Transmission model**
To simulate SARS-CoV-2 transmission, we used a variant of the SLIR (susceptible, latent, infectious, and removed) model, where infectious individuals are further classified as pre-symptomatic (P), symptomatic (I), and asymptomatic (A). If a susceptible individual $i$ has contact with an infectious individual $j$, the susceptible individual has a certain layer-specific risk of acquiring the infection.

The transmission parameters in the model are set according to the literature on the ancestral SARS-CoV-2 lineage. The incubation period follows a gamma distribution with a mean of 6.3 days and a standard deviation of 4.3 (shape=2.08, rate=0.33) (44). We consider transmission to start 2 days before symptom onset (44). We also consider an age-specific susceptibility to infection according to the literature (45). The duration of the infectious period was chosen such that the generation time distribution was equal to 7.0 days as reported in (46). We calibrated the model to match the fraction of infections reported in the four layers in China before the lockdown (35). All parameters used in the model can be found in Additional File 1.

**Alternative implementations of TTI**

TTI is a symptom-based testing, case-based tracing, and isolation process that will be modelled as follows:

**Symptom-based testing:** Symptomatic individuals undergo a reverse transcription polymerase chain reaction (RT-PCR) test with a given probability of testing positive ($P_{test}$). We denote the delay between the symptom onset and sample collection as $T_{st}$ and the delay from sample collection to laboratory diagnosis as $T_{tr}$. The laboratory diagnosis result considers the sensitivity of the RT-PCR test based on the timing of sample collection (47). Individuals' contact patterns do not change during the delay window between the onset of symptoms and the receipt of the RT-PCR test.

**Contact tracing:** If the result of the RT-PCR test is positive, the contact tracing process is initiated. In this study, we trace both the primary contacts of the infected individual case and the contacts of those primary contacts (secondary contacts). Specifically, all household members of the infected individual are traced. School and/or workplace contacts have a given probability of being traced ($P_{trace}$) if they had contact with the infected individual within a time window of length $T_{trace}$ surrounding the date the infected individual took the RT-PCR test. As most community contacts occur in public places and are difficult to trace in the real world, we assumed that community layer contacts could not be traced.

**Isolation:** The infected individual is isolated at home for 14 days. Their traced contacts are quarantined at home (48) immediately, and they undergo RT-PCR testing. While quarantined in their place of residence, individuals have contacts only in the household layer. If the traced contact receives a negative test result, they resume their regular contact patterns. Otherwise, they isolate at home for 14 days.

After symptom onset and the receipt of a positive RT-PCR test, we model the following alternative implementations of TTI for the infected individual:

1) **TTI-L1**: Isolation at home for 14 days without contact tracing;
2) **TTI-L2**: Isolation at home for 14 days where primary contacts are traced and required to take an RT-PCR test;
3) **TTI-L3**: Isolation at home for 14 days where primary contacts are traced and isolated at home for 14 days without testing;
4) **TTI-L4:** Quarantine at a dedicated facility for 14 days where primary contacts are traced and quarantined at dedicated facilities for 14 days without testing;
5) **TTI-L5**: Isolation at home for 14 days where both the primary and secondary contacts are traced and required to take an RT-PCR test;
6) **TTI-L6**: Isolation at home for 14 days where both the primary and secondary contacts are

traced and isolated at home for 14 days directly without testing;

7) **TTI-L7**: Quarantine at a dedicated facility for 14 days where both the primary and secondary contacts are traced and quarantined at dedicated facilities for 14 days without testing.

We are using L5 as our baseline implementation of TTI. The alternative implementations of TTI are shown in Fig. 2

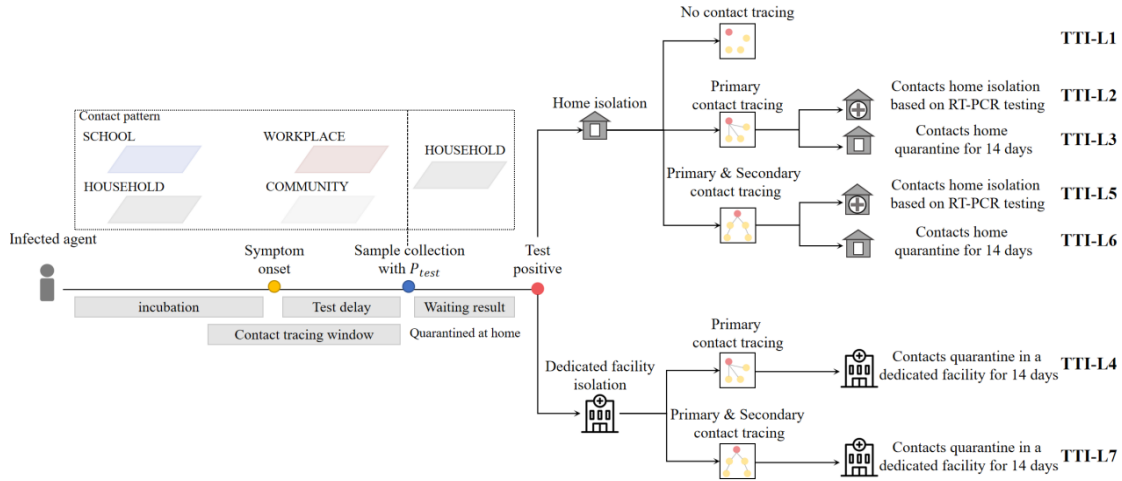

**Fig. 2 Schematic representation of the alternative implementations of TTI.** Once an individual is infected, they enter an incubation period, followed by the onset of symptoms. After symptom onset, testing is conducted with the possibility of $P_{test}$, often after a delay, with an additional waiting period for the results. During this waiting period, the individual is quarantined at home. If the test result is positive, contact tracing is carried out retrospectively over a specified tracing window. Identified contacts are then managed according to different TTI strategies, ranging from no contact tracing (TTI-L1) to primary and secondary contact tracing with facility-based quarantine (TTI-L7). Each TTI implementation level (TTI-L1 to TTI-L7) defines specific actions for contacts, such as home isolation based on RT-PCR testing, home quarantine, or facility quarantine.

**Other NPIs**

In conjunction with TTI, here we propose two other NPIs that were commonly adopted in the early phase of the COVID-19 pandemic:

1) **Social distancing**, which is modeled by limiting random contacts by reducing a fraction of contacts for each individual in the community layer;

2) **Partial closure of school and workplace**, which is modeled by removing all contact within the school layer and removing contact in 10%, 30% and 50% of workplaces. Moreover, we considered a 50% reduction of contacts in the community layer.

Noting that these two NPIs are implemented all the time. Details about NPIs are reported in Additional File 1.

**Metrics used for the evaluation of the effectiveness of an intervention**

We adopt the Epidemic Prevention Potential (EPP) (49) to measure the impact of TTI in containing an outbreak. The epidemic prevention potential is defined as $EPP = 1 - P_I/P_{NI}$ where $P_I$ and $P_{NI}$ represent the probabilities of an uncontained outbreak when an intervention is deployed (e.g., TTI) and when no interventions are deployed, respectively. Here, we consider an outbreak if the cumulative number of symptomatic cases in the simulation reaches 400 within 90 days. As the local transmission evolves into an epidemic outbreak, we assessed the effectiveness of the analyzed intervention in mitigating an outbreak through the following metrics: reduction in peak daily incidence of new symptomatic infections, peak timing, and final symptomatic infection attack rate. Moreover, we estimated the "costs" associated with the deployed intervention in terms of mean number of days spent in isolation/quarantine per person and the maximum number of individuals simultaneously isolated/quarantined.

We explored reproduction numbers within the range of 1.3 to 3.1 to represent the different transmissibility scenarios. To evaluate the likelihood of the strategy in controlling an outbreak, we initialized the model with one infected individual.

**Results**
**Baseline scenario**
By assuming the baseline value of model parameters, the probability of an uncontained outbreak in the "no intervention" scenario increases from about 6.1% for $R_0$=1.3 to 90.1% for $R_0$=3.1 (Fig. 3A). The peak daily incidence of new symptomatic infections increases from 0.75 per 1000 individuals (95% CI: 0.69-0.83) for $R_0$=1.3 to 10.5 (95% CI: 10.3-10.8) per 1000 individuals for $R_0$=3.1 (Fig. 3B). The epidemic peak time ranges from 268 days (95%CI: 227-324 days) for $R_0$=1.3 to 59 days (95%CI: 55-64 days) for $R_0$=3.1 (Fig. 3C). The final symptomatic infection attack rate ranges from 8.1% (95%CI: 6.7%-8.6%) for $R_0$=1.3 to 23.5% (95%CI: 23.4%-23.6%) for $R_0$=3.1 (Fig. 3D).

When considering TTI-L5, the EPP decreases from 88.2% (95% CI: 87.9%-88.5%) for $R_0$=1.3 to 2.5% (95%CI: 0.9%-4%) for $R_0$=1.9, and it becomes negligible for higher values of the reproduction number (Fig. 3A). The peak daily incidence of new symptomatic infections is reduced by 76.1% (95% CI: 71.6%-82.9%) for $R_0$=1.3 and 64.0% (95% CI: 63.7%-64.3%) for $R_0$=3.1 (Fig. 3B). The epidemic peak is delayed by 55 days for $R_0$=1.3 and 6 days for $R_0$=3.1 (Fig. 3C). The reduction in symptomatic attack rate ranges from 74.3% (95% CI: 62.6%-93.5%) for $R_0$=1.3 to 18.8% (95% CI: 18.1%-19.4%) for $R_0$=3.1 (Fig. 3D). The implementation of this intervention entails costs in terms of isolated/quarantined individuals. The mean isolation/quarantine period per individual increases from 11 days (95%CI: 3-15) for $R_0$=1.3 to 43 days (95%CI: 42-44) for $R_0$=1.9, and it then stabilizes around 40 days for $R_0$ >1.9 (Fig. 3E). The peak daily number of simultaneously isolated/quarantined individuals per 1000 individuals increases with the reproduction number and reaches 507 (95%CI: 500-515) when $R_0$=3.1 (Fig. 3F).

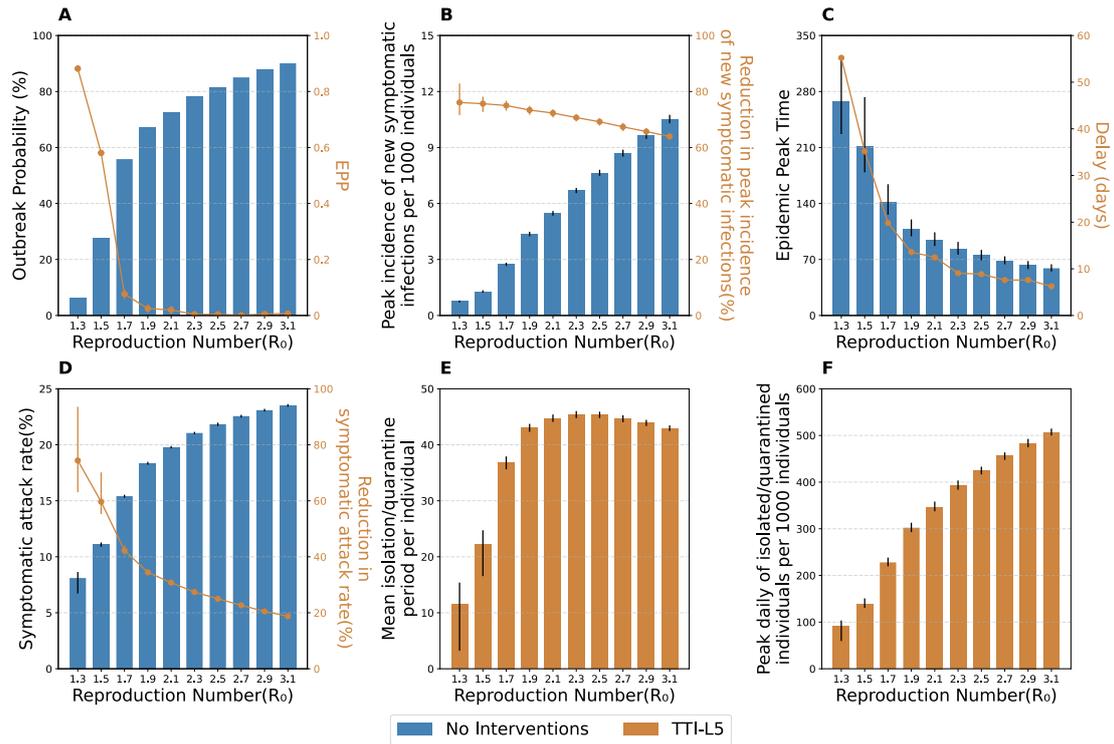

**Fig. 3 Effectiveness and implementation cost of TTI in the baseline scenario. A** Probability of an outbreak without interventions and the Epidemic Prevention Potential (EPP) under TTI-L5 across various reproduction numbers. **B** Peak incidence of new symptomatic infections per 1,000 individuals and corresponding reduction under TTI-L5 for different $R_0$ values. **C** Epidemic peak time and delay in peak time achieved with TTI-L5 for different $R_0$ values. **D** Symptomatic attack rate and corresponding reduction achieved with TTI-L5 for different $R_0$ values. **E** Mean time spent in isolation or quarantine per individual under TTI-L5. **F** Peak daily number of isolated or quarantined individuals per 1,000 individuals under TTI-L5. In all panels, the vertical error bars indicate the 95% CI.

**Effect of the network structure and assumptions on TTI implementation**

We conducted various sensitivity analyses by varying parameters related to the network structure (i.e., static vs. temporal network, different fractions of recurring contacts) and the parameters regulating the implementation of TTI: the delay between symptom onset to sample collection, the delay between sample collection and the test result, the probability of testing a symptomatic individual, the probability of being traced, and the length of the tracing time window. In the main text (Fig. 4), we report only a summary of these sensitivity analyses; the complete results are provided in Additional File 1.

For the TTI-L5 strategy, our results show that considering a static contact network has little effect on the estimated EPP (Fig. 4A). However, we found remarkably lower reductions in the peak daily incidence of new symptomatic infections (Fig. 4B) and on the final symptomatic infections attack rate (Fig. 4C). Shortening the time delay from symptom onset to sample collection can effectively increase the EPP when the reproduction number is lower than 1.9, while it does not have a remarkable effect for higher values of $R_0$ (Fig. 4D). Considering a shorter delay has limited effect on the peak daily incidence of symptomatic infections and on the final symptomatic attack rate (Fig. 4E-F). On the other hand, the duration of the contact tracing time window has a larger impact on the peak daily incidence of symptomatic infections and final symptomatic attack rate than on the EPP (Figs. 4G-I). Moreover, our sensitivity analyses show that the fraction of recurring contacts, delay in test result, testing probability, and tracing probability have limited effect on the estimated impact of TTI-L5 (Figs. 6, 8-10 in Additional File 1).

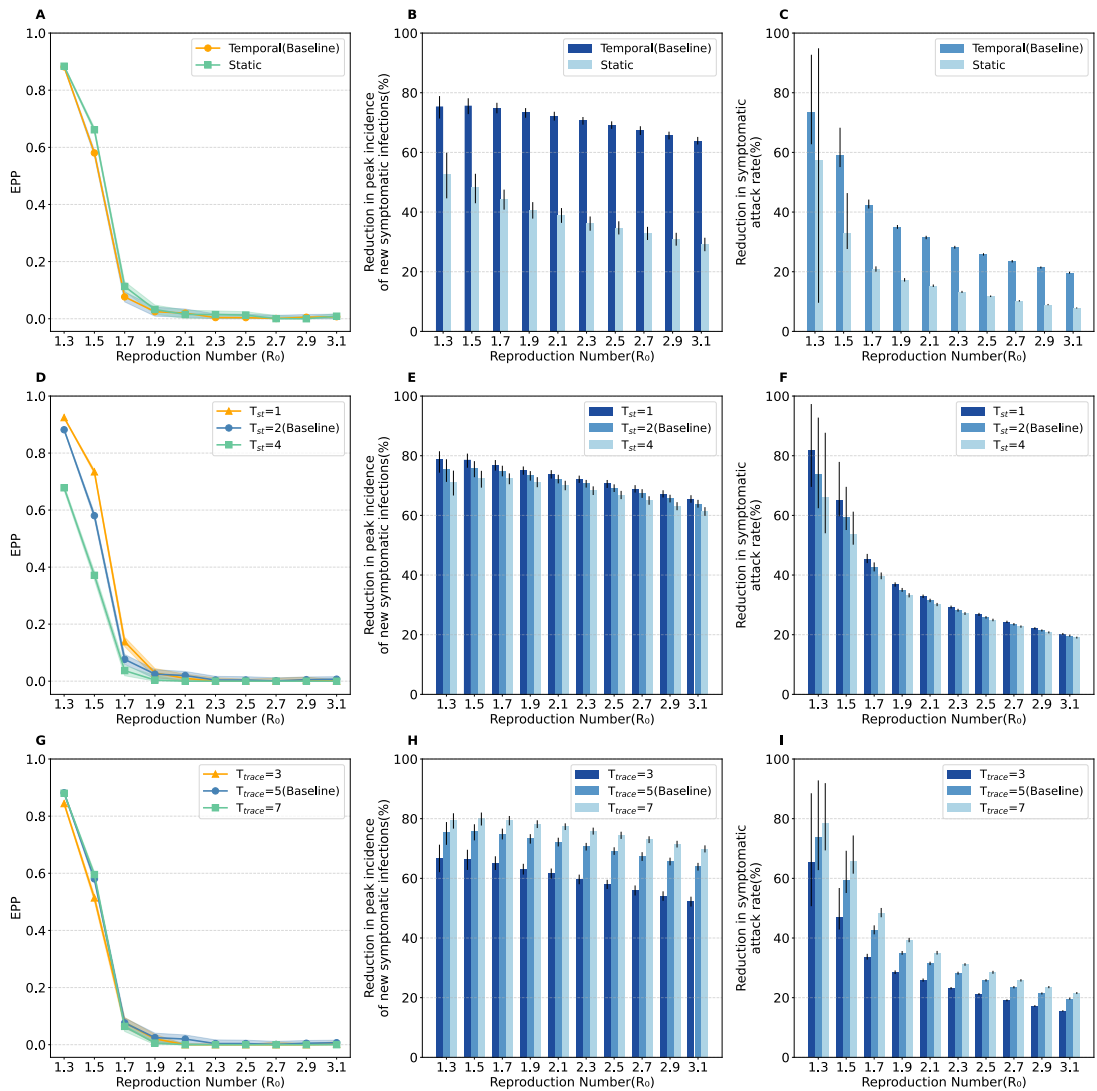

**Fig. 4 Effectiveness of sensitivity experiments.** **A** Epidemic Prevention Potential (EPP) of the TTI strategy under temporal and static contact networks. Shaded areas represent the 95% confidence interval (CI). **B** Reduction in peak incidence of new symptomatic infections under temporal and static contact networks, with vertical error bars indicating the 95% CI. **C** Reduction in symptomatic attack rate under temporal and static contact networks, with vertical error bars indicating the 95% CI. **D** as in **A** but for different delays from symptom onset to RT-PCR testing. **E** As in **B** but for different delays from symptom onset to RT-PCR testing. **F** as in **C** but for different delays from symptom onset to RT-PCR testing. **G** as in **A** but for different tracing windows. **H** as in **B** but for different tracing windows. **I** as in **C** but for different tracing windows.

## Comparison between different TTI implementations

Our results show that strategies considering quarantine of contacts (TTI-L3, TTI-L4, TTI-L6, TTI-L7) have a larger EPP than strategies that either avoid tracing contacts or require contacts to be tested before isolation (TTI-L1, TTI-L2, TTI-L5) (Fig. 5A). Moreover, strategies that trace both primary and secondary contacts perform particularly well (Fig. 5A).

For uncontained epidemics, we found that the peak daily incidence of new symptomatic infections and the final symptomatic attack rate decreases as the level of TTI implementation increases (L1 to L7) (Figs. 5 B,D). In particular, we observe a large difference between TTI-L1,2 (no contact tracing and home isolation of primary contacts based on testing, respectively) and TTI-L3 (home quarantine of primary contacts) (Figs. 5B,D). In terms of peak timing, the pattern is less clear as the level of TTI increases, the incidence becomes flatter and with high level of TTIs we can observe either a highly delayed epidemic peak or a small delay (or even an anticipated peak) epidemic peak of a remarkably reduced size (Figs. 5B,C). Moreover, we found that given a value of the reproduction number, in general the cost associate with the intervention increases as the intervention level increases (Figs. 5E,F). A notable exception is represented by TTI-L6 (home quarantine of primary and secondary contacts), which is less effective than TTI-L7 (quarantine of primary and secondary contacts in dedicated facilities) in mitigating an outbreak, but entails a larger number of days in isolation/quarantine than it.

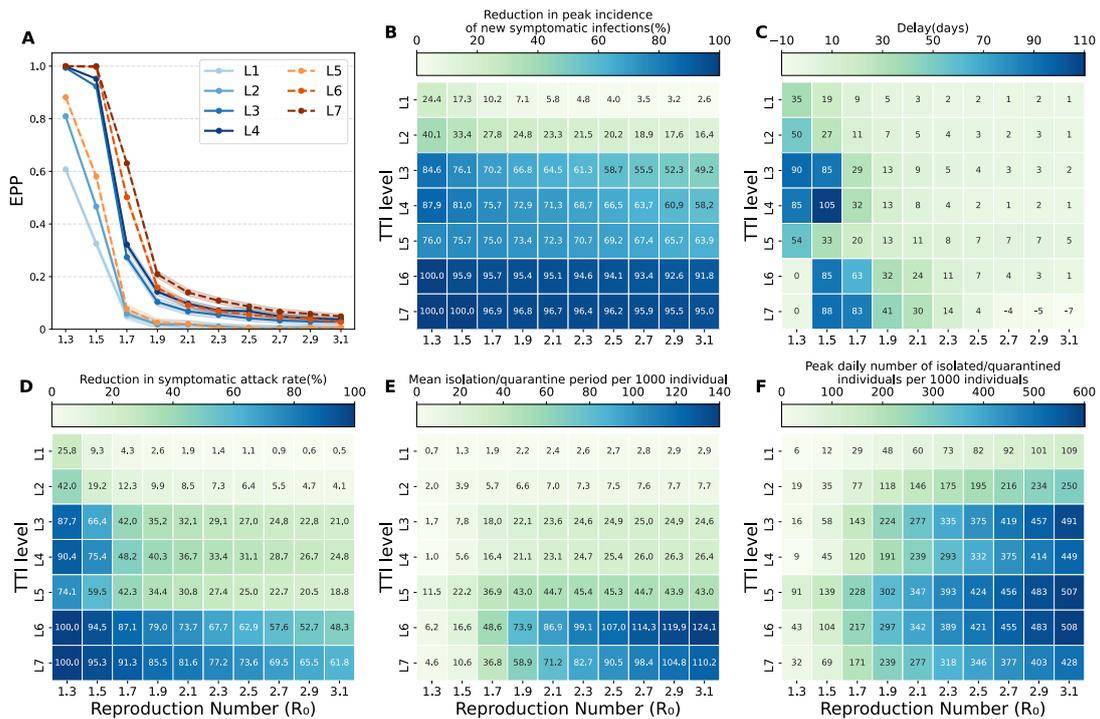

**Fig. 5 Effectiveness of different implementation processes. A** EPP for TTI across different TTI levels and $R_0$ values. Dashed lines represent tracing of both direct and secondary contact tracing, while solid lines represent tracing of direct contact tracing only. **B** Reduction in peak incidence of new symptomatic infections across different TTI levels and $R_0$ values. **C** Delay in the peak timing across different TTI levels and $R_0$ values. **D** Reduction in symptomatic attack rate across different TTI levels and $R_0$ values. **E** Mean isolation or quarantine period per 1,000 individuals across different TTI levels and $R_0$ values. **F** Peak daily number of isolated or quarantined individuals per 1,000 individuals across different TTI levels and $R_0$ values.

**Social distancing scenario and enhanced scenario**

Drawing from NPIs implemented during the COVID-19 pandemic, we integrated TTI with social distancing measures targeting community contacts, such as those occurring during public events and leisure activities. The combination of TTI with a limitation of community contacts significantly enhanced the effectiveness of the intervention across all the analyzed $R_0$ values, increasing both the likelihood of containing an outbreak (Fig. 6A) or, should an outbreak unfold, mitigating its burden (Fig. 6B), but entail high social costs (Fig. 6C). Furthermore, we analyzed a scenario assuming a 50% reduction of community contacts, fully online education, and partial reduction of in-person work. This combined strategy provides higher likelihoods of successfully containing an outbreak, although the EPP is lower than 50% for $R_0>2$ (Fig. 6D). The mitigation effect of this strategy is estimated to be above 35% for $R_0$ up 3.1 (Fig. 6E), although the prolonged application of this strategy would lead to major social costs (Fig. 6F). Compared to Fig. 6C, the reduction in the mean isolation/quarantine period per individual in Fig. 6F is attributed to the enhanced effectiveness of the intervention resulting from the closure of schools and workplaces, but at the cost of hindering normal operations in both educational and professional settings.

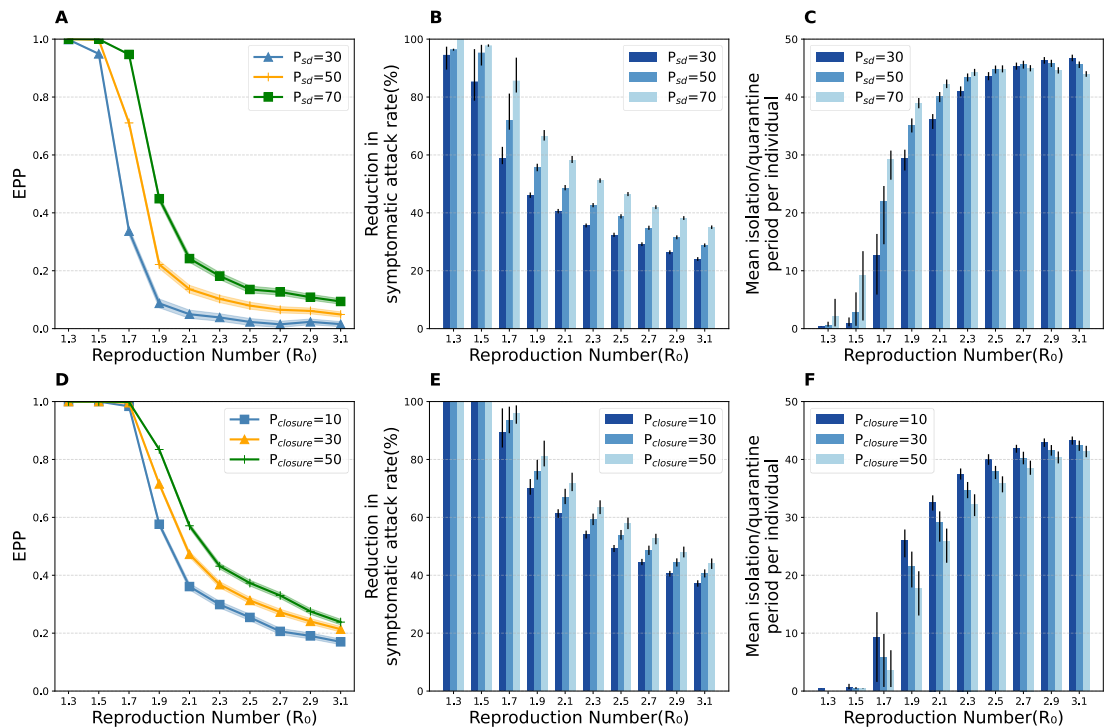

**Fig. 6 Effectiveness of TTI in combination with other social distancing measures. A** EPP across various reproduction numbers for TTI-L5 in combination with a 30%, 50%, or 70% reduction in the number of community contacts. **B** As A but for the reduction in the symptomatic attack rate. **C** As A but for the mean number of days spent in isolation or quarantine per individual. **D** EPP across various reproduction numbers for TTI-L5 in combination with a 50% reduction of community contacts, fully online education, and 10%, 30%, or 50% reduction of in-person contacts at the workplace. **E** as D but for the reduction in the symptomatic attack rate. **F** as D but for the mean number of days spent in isolation or quarantine per individual.

**Discussion**

In this study, we developed a multi-layer synthetic population encompassing households, schools,

workplaces, and community based on public records. To model the complex contact patterns of real-world contacts, we then generated a temporal contact network determining contacts between agents in the synthetic population based on diary-based contact survey data collected in 2017-18 in Shanghai, China. This multi-layer temporal network was used to simulate the spread of respiratory pathogen and perform a model-based evaluation of the effectiveness of alternative implementations of TTI, using SARS-CoV-2 as a case study. Our results indicate that TTI can be effective in containing an outbreak for values of the reproduction number up to 1.3. Should an outbreak start to unfold, TTI can mitigate disease burden but it would entail considerable social costs. Our results show that strategies considering quarantine of contacts lead to a larger EPP than strategies that either avoid tracing contacts or require contacts to be tested before isolation. Moreover, strategies that trace both primary and secondary contacts perform particularly well. Combining TTI with other social distancing measures can improve the likelihood of successfully containing an outbreak but the estimated EPP remains lower than 50% for reproduction numbers higher than 2.1.

Our model offers several advantages as compared to theoretical network-based models. First, the synthetic population closely fits socio-demographic data. Second, we developed a multi-layer temporal contact network based on real contact survey data, where each layer in the synthetic population has distinct connection rules and contact number distributions. Moreover, the contact weights for each layer were calibrated based on epidemiological data from the literature, enabling a more accurate simulation of complex contact patterns in real environments. We conducted sensitivity analyses by varying parameters related to the network structure (e.g., static vs. temporal networks, contact heterogeneity). Our results indicate that temporal networks with uncorrelated, heterogeneous contacts lead to a higher peak of infections and a larger attack rate compared to static networks. However, they also show a higher effectiveness of TTI. This highlights the importance of understanding the contact network and its temporal variations to properly assess epidemic spread and the effectiveness of TTI. New studies are being conducted to estimate the "new normal" for social contacts after the COVID-19 pandemic (50) and to understand the temporal variations in the number of contacts (51), but new knowledge is also warranted to quantify the frequency and likelihood of repeated interactions.

Our analysis has several limitations. First, we made several optimistic estimates regarding the implementation of intervention measures, such as assuming zero delay from case confirmation to initiating contact tracing, unlimited PCR capacity and immediate implementation of social distancing measures. In reality, all these parameters should be tailored to the real-world conditions and capacity of the study site. Previous research has already demonstrated that these factors significantly affect the effectiveness of TTI. Second, our assessment of the costs associated with TTI is entirely based on the number of days spent in isolation/quarantine. Other indicators should be considered to estimate the cost of the intervention more broadly (e.g., economic costs, mental health, loss of productivity for care givers). Third, we assumed that all individuals in the community layer were potentially exposed to the pathogen; this represents a coarse-grained assumption that could be improved when new data on social interactions will be available. Additionally, the contact patterns data that we used to calibrate the model was collected before the COVID-19 epidemic, and thus the extent to which this data is representative of the "new normal" remains unclear.

In conclusion, our model-based evaluation highlights the challenges of relying on TTIs to contain an outbreak of a novel pathogen with characteristics similar to SARS-CoV-2. Our analysis also shows that the estimated effectiveness of TTI depend on the way contact patterns are modeled, supporting the relevance of obtaining updated and comprehensive data on human social interactions.

**List of abbreviations**
COVID-19: Coronavirus disease 2019
SARS-CoV-2: severe acute respiratory syndrome coronavirus 2
TTI: test-trace-isolate
EPP: Epidemic Prevention Potential
RT-PCR: Reverse Transcription-Polymerase Chain Reaction
NPI: Nonpharmaceutical interventions

**Declarations**
**Ethics approval and consent to participate**
Not applicable

**Consent for publication**
Not applicable

**Availability of data and materials**
The synthetic population is available from the corresponding author on reasonable request. Other data and codes used in the study are provided in Supplementary Information and are available on GitHub at https://github.com/Cyyloud/Test-Trace-Isolate-Model

**Competing interests**
The authors declare that they have no competing interests.

**Funding**
This work was supported by the National Natural Science Foundation of China (No. 62373264), the Major Program of National Fund of Philosophy and Social Science of China (No. 20&ZD112), and the 111 Project under grant agreement B21044.

**Authors' contributions**
Q-HL and MA designed the study. Q-HL supervised the study. YC carried out development of the model, designed the simulations, prepared the figures, and performed the statistical analysis. Q-HL, MA, and YC drafted the manuscript. RW, WW, LY helped with developing the code for model, constructing the figures and commenting on the manuscript. GQS, JL, AGK, PCV, Q-HL made critical revision of the manuscript for important intellectual content. All authors were involved in the research and revising of the manuscript. All authors read and approved the final manuscript.

**Acknowledgements**

Not applicable

**Supporting information**
Additional File 1

variations in human contact patterns and their impact on the transmission of respiratory infectious diseases. Influenza and Other Respiratory Viruses. 2024;18(5):e13301.